# Two stellar components in the halo of the Milky Way

Daniela Carollo[1,2,3,5], Timothy C. Beers[2,3], Young Sun Lee[2,3], Masashi Chiba[4], John E. Norris[5] , Ronald Wilhelm[6], Thirupathi Sivarani[2,3], Brian Marsteller[2,3], Jeffrey A. Munn[7], Coryn A. L. Bailer-Jones[8], Paola Re Fiorentin[8,9], & Donald G. York[10,11]

*[1]INAF - Osservatorio Astronomico di Torino, 10025 Pino Torinese, Italy, [2]Department of Physics & Astronomy, Center for the Study of Cosmic Evolution, [3]Joint Institute for Nuclear Astrophysics, Michigan State University, E. Lansing, MI 48824, USA, [4]Astronomical Institute, Tohoku University, Sendai 980-8578, Japan, [5]Research School of Astronomy & Astrophysics, The Australian National University, Mount Stromlo Observatory, Cotter Road, Weston Australian Capital Territory 2611, Australia, [6]Department of Physics, Texas Tech University, Lubbock, TX 79409, USA, [7]US Naval Observatory, P.O. Box 1149, Flagstaff, AZ 86002, USA, [8]Max-Planck-Institute für Astronomy, Königstuhl 17, D-69117, Heidelberg, Germany, [9]Department of Physics, University of Ljubljana, Jadronska 19, 1000, Ljubljana, Slovenia, [10]Department of Astronomy and Astrophysics, Center, [11]The Enrico Fermi Institute, University of Chicago, Chicago, IL, 60637, USA*

**The halo of the Milky Way provides unique elemental abundance and kinematic information on the first objects to form in the Universe, which can be used to tightly constrain models of galaxy formation and evolution. Although the halo was once considered a single component, evidence for is dichotomy has slowly emerged in recent years from inspection of small samples of halo objects. Here we show that the halo is indeed clearly divisible into two broadly overlapping structural components -- an inner and an outer halo – that exhibit different spatial density profiles, stellar orbits and stellar metallicities (abundances of elements heavier than helium). The inner halo has a modest net prograde rotation, whereas the outer halo exhibits a net retrograde rotation and a peak metallicity one-third that of the inner halo. These properties indicate that the individual halo components probably formed in fundamentally different ways,**

**through successive dissipational (inner) and dissipationless (outer) mergers and tidal disruption of proto-Galactic clumps.**

Astronomers have long sought to constrain models for the formation and evolution of the Milky Way (our Galaxy) on the basis of observations of the stellar and globular cluster populations that it contains. These populations are traditionally defined as samples of objects that exhibit common spatial distributions, kinematics and metallicities (the age of a population, when available, is also sometimes used). Metallicity is taken by astronomers to represent the abundances of the elements heavier than helium, which are only created by nucleosynthesis in stars – either internally via nuclear burning in their cores or externally during explosive nucleosynthesis at the end of their lives. The earliest generations of stars have the lowest metallicities, because the gas from which they formed had not been enriched in heavy elements created by previous stars and distributed throughout the primordial interstellar medium by stellar winds and supernovae.

Previous work has provided evidence that the halo of the Milky Way may not comprise a single population, primarily from analysis of the spatial profiles (or inferred spatial profiles) of halo objects [1--4]. A recent example of such an analysis is the observation of two different spatial density profiles for distinct classes of RR Lyrae variable stars in the halo[5]. In addition, tentative claims for a net retrograde motion of halo objects by previous authors supports the existence of a likely dual-component halo [6--10]. The central difficulty in establishing with confidence whether or not a dichotomy of the halo populations exists is that the past samples of tracer objects have been quite small, and usually suitable only for consideration of a limited number of the expected signatures of its presence.

In the present work, we examine this question in detail using a large, homogeneously selected and analysed sample of over 20,000 stars, originally obtained as calibration data during the course of Sloan Digital Sky Survey (SDSS)[11]. Although there are many possible alternative (and more complex) models that might be considered, multiple lines of evidence derived from these data clearly confirm that the halo can be resolved into (at least) two primary populations, the inner and the outer halo, with very different observed properties.

We find that the inner-halo component of the Milky Way dominates the population of halo stars found at distances up to 10-15 kpc from the Galactic Centre (including the Solar neighbourhood). An outer-halo component dominates in the regions beyond 15-20 kpc. We show the inner halo to be a population of stars that are non-spherically distributed about the centre of the Galaxy, with an inferred axial ratio on the order of ~ 0.6. Inner-halo stars possess generally high orbital eccentricities, and exhibit a modest prograde rotation (between 0 and 50 km $s^{-1}$) around the centre of the Galaxy (see Supplementary Table 1). The distribution of metallicities for stars in the inner halo peaks at [Fe/H] = −1.6, with tails extending to higher and lower metallicities. (Here, metallicity is defined as $[A/B] = \log_{10} (N_A/N_B) - \log_{10} (N_A/N_B)_{\odot}$ , where $N_A$ and $N_B$ represent the number density of atoms of elements A and B, and the subscript ⊙ indicates solar values.) The outer halo, by contrast, comprises stars that exhibit a much more spherical spatial distribution, with an axial ratio ~ 0.9 to 1.0. Outer-halo stars cover a wide range of orbital eccentricities, including many with lower eccentricity orbits than found for most stars associated with the inner halo, and exhibit a clear retrograde net rotation (between –40 and –70 km $s^{-1}$) about the centre of the Galaxy. The metallicity distribution function (MDF) of the outer halo peaks at lower metallicity than that of the inner halo, around [Fe/H] = −2.2, and includes a larger fraction of low-metallicity stars than does the MDF of the inner-halo population.

## Evidence for the dichotomy of the halo

The spectroscopy, photometry, and astrometry for our large sample of stars were obtained from observations carried out with the Apache Point 2.5-m SDSS telescope; these data are publicly available as Data Release 5[12]. Details concerning the selection of the stars and the measurement of their stellar parameters (temperature, surface gravity, and metallicity, [Fe/H]), as well as the methods used to obtain their estimated distances, proper motions and derived kinematics, can be found in the Supplementary Information.

Our 20,236 program stars explore distances up to 20 kpc from the Sun, but we can only obtain useful estimates of the full space motions (as described in the Supplementary Information) for the subset of 10,123 stars in a local volume (up to 4 kpc from the Sun; Fig. 1). The restriction of the sample to the region of the solar neighbourhood is also made so that the assumptions going into the kinematic calculations are best satisfied. Figure 2 shows the distribution of [Fe/H] for different cuts in the V velocity, which is the orbital component that measures the motion of a star (with respect to the Local Standard of Rest) in the rotation direction of the Galaxy. The transition in the distribution of [Fe/H] that is expected as one sweeps from stars with thick-disk-like motions to stars with halo-like motions is clear. However, with the large sample of stars in our sample, it is possible to investigate the change in the distribution of [Fe/H] for stars that are increasingly more retrograde, as well as for those that are both highly retrograde and have orbits taking them to high $Z_{max}$ (the maximum distance above the Galactic plane reached by a star during the course of its orbit about the Galactic Centre – the Supplementary Information describes the methods used to derive this fundamental parameter). This figure shows that stars with the most retrograde orbits, and those that reach large distances in their orbits above the Galactic plane, exhibit distributions of [Fe/H] that peak at metallicities between −2.0 and −2.2, which we associate with the outer-halo population. The inner-halo population dominates the samples of stars with peak metallicity [Fe/H] ≈ −1.6.

Astronomers have long debated whether there might exist a change in the rotational properties of the halo of the Milky Way as a function of distance from the Galactic Centre, based on much smaller samples of globular clusters[2,10] and stars[6--9,13,14] than we consider here. The stellar samples were obtained with selection criteria (for example, on the basis of high proper motions for halo stars in the solar neighbourhood[7,13,14], or from *in situ,* apparent-magnitude limited surveys[6,9,15]) that we suggest favoured membership in one or the other of the now-clearly-revealed halo components. A detailed summary of the kinematics of our programme stars is presented in the Supplementary Information, where we also establish consistency between properties obtained through techniques based on full space motions and those based on radial velocities alone, which argues against the existence of any large systematic errors in the proper motions.

Table 1 summarizes the past and present determinations of $<V_{\Phi}>$, the mean rotational velocity with respect to the Galactic Centre, where claims for a retrograde halo have been made. Previous samples that have addressed this question were based on either much smaller total numbers of objects (with the limitation that they could not well sample both the inner- and outer-halo populations), did not have proper motions available (or only highly uncertain ones), or were otherwise restricted due to the selection criteria employed (that is, they were kinematically biased[16], or had limited sky coverage, rendering them sensitive to the effects of individual star streams[17]. The local sample of SDSS calibration stars we have assembled does not suffer from any of these limitations. The retrograde signatures for stars we associate with the outer-halo population are robust and highly statistically significant (except for the smallest subsample). However, even our precise present determination of the net retrograde rotation of the outer halo, based on our local sample, is probably influenced by some degree of overlap between outer-halo stars with those from the inner-halo population.

The distribution of [Fe/H] for stars on increasingly retrograde orbits about the Galactic Centre for subsamples that reach different distances from the Galactic plane ($Z_{max}$) is shown in Fig. 3. The MDFs of the stars with $Z_{max}$ close to the Galactic plane are very different from those whose orbits reach farther from the plane. The distribution of metallicity clearly shifts to lower abundances as more severe cuts on $V_{\Phi}$ or $Z_{max}$ are applied, as supported by rigorous statistical tests. We conclude that the halo of the Galaxy comprises stars with intrinsically different distributions of [Fe/H]; the observed changes in the MDF of halo stars with $V_{\Phi}$ and $Z_{max}$ would not be expected if the halo is considered as a single entity.

The Supplementary Information presents additional observed differences in the energetics, the distribution of orbital eccentricities, and changes in the nature of stellar orbits for our programme stars that are also inconsistent with a single halo population.

In order to provide confirmation of the shift in the MDF inferred from our analysis of a local sample of stars, we also examine an auxiliary sample of stars that are presently located much farther from the Galactic Centre. This sample comprises 1235 blue horizontal-branch stars selected from the SDSS[18]. The stars cover a wide range of distances, from 5 kpc to over 80 kpc from the centre of the Galaxy. Statistical tests strongly reject the hypothesis that the stars at large distances from the Galactic centre could be drawn from the same parent population as those at distances close to the Galactic Centre (Fig. 4).

It has been shown, on the basis of Jeans' theorem[19,20], that the global structure of the stellar halo can be recovered from local kinematic information, as long as one has a sufficiently large number of stars observed in the solar neighbourhood that explore the full phase-space distribution of the pertinent stellar populations. We note that the actual halo systems of the Galaxy are unlikely to be in well-mixed equilibrium states. However, the relaxation process is very slow compared to the orbital periods of typical stars, so the Jeans' theorem and the approach based on it remain at least approximately valid. The result of this exercise for our large sample of SDSS calibration stars, over narrow cuts in metallicity, is shown in Fig. 5. The observed changes in the inferred spatial density profiles suggest that a flattened inner-halo population dominates locally for stars with [Fe/H] > −2, whereas the outer-halo population has a nearly spherical distribution, and dominates at distances beyond $r \approx 15$-20 kpc (where r represents the distance from the Galactic Centre), as well as locally for stars with Fe/H] < −2.0. Variations in the halo spatial profile with distance have been recognized by a number of previous authors[1--4,15,20], based on samples of stars that are one to two orders of magnitude smaller than our present data set.

**Implications of the dichotomy of the halo**

An early model for the formation of the Galaxy, based on the rapid (a few hundred million years) monolithic collapse of a gaseous proto-Galaxy[21], has yielded to the more recent idea that the halo of the Galaxy was assembled, over the span of several billion years, from smaller proto-Galactic clumps[22]. This hierarchical assembly model has received close attention in recent years, in part because it fits well with the prevailing theory for the formation and evolution of structure in the Universe, based on the early collapse of 'mini-haloes' of cold dark matter (CDM)[23,24]. Modern numerical simulations for the assembly of large spirals based on CDM cosmogonies predict that the stars in the haloes of galaxies like the Milky Way might be comprised of the shredded stellar debris of numerous dwarf-like galaxies that have been torn apart by tidal interactions with their parent galaxy[24--27]. Recent quantitative analysis of the amount of structure visible in the halo of the Galaxy from SDSS[28--31] imaging provides compelling additional evidence[32]. Others have argued that some combination of a monolithic collapse and a hierarchical assembly model may be necessary to fully explain the observed data[2,15,33].

Within the context of the CDM model, the formation of the inner halo may be understood in the following manner. Low-mass sub-Galactic fragments are formed at an early stage. These fragments rapidly merge into several (in many simulations, two[26,34]) more-massive clumps, which themselves eventually dissipatively merge (due to the presence of gas that has yet to form stars). The essentially radial merger of the few resulting massive clumps gives rise to the dominance of the high-eccentricity orbits for stars that we assign here to membership in the inner halo. Star formation within these massive clumps (both pre- and post-merger) would drive the mean metallicity to higher abundances. This is followed by a stage of adiabatic compression (flattening) of the inner halo component owing to the growth of a massive disk, along with the continued accretion of gas onto the Galaxy[34,35].

The fact that the outer-halo component of the Milky Way exhibits a net retrograde rotation (and a different distribution of overall orbital properties), as found here, clearly indicates that the formation of the outer halo is distinct from that of both the inner-halo and disk components. We suggest, as others have before, that the outer-halo component formed, not through a dissipative, angular-momentum-conserving contraction, but rather through dissipationless chaotic merging of smaller subsystems within a pre-existing dark-matter halo. These subsystems would be expected to be of much lower mass, and subject to tidal disruption in the outer part of a dark-matter halo, before they fall farther into the inner part. As candidate (surviving) counterparts for such subsystems, one might consider the low-luminosity dwarf spheroidal galaxies surrounding the Galaxy, in particular the most extreme cases recently identified from the SDSS[36,37]. Subsystems of lower mass, and by inference, even lower metallicity, may indeed be destroyed so effectively that none (or very few) have survived to the present day. If so, the outer-halo population may be assembled from relatively more metal-poor stars, following the luminosity-metallicity relationship for Local Group dwarf galaxies[38]. The net retrograde rotation of the outer halo may be understood in the context of the higher efficiency of phase mixing for the orbits of stars that are stripped from subsystems on prograde, rather than retrograde orbits[39,40].

The clear difference in the MDFs of the two halo populations we identify also suggests that the lowest metallicity stars in the Galaxy may be associated with the outer halo, which can be exploited for future directed searches. It is noteworthy that the hyper metal-poor stars HE 0107-5240[41] and HE 1327-2326[42], both of which have [Fe/H] < −5.0, as well as the recently discovered ultra metal-poor star HE 0557-4840[43], with [Fe/H] = −4.8, are either located greater than 10 kpc away (HE 0107-5240, HE 0557-4840) or have space motions that carry them far out into the Galaxy (HE 1327-2326; A. Frebel, personal communication).

In addition, efforts to determine the primordial lithium abundance from observations of the most metal-poor stars[44] may have inadvertently mixed samples from the inner- and outer-halo populations; such stars could have formed and evolved in rather different astrophysical environments. The inner/outer halo dichotomy may also have an impact of the expected numbers of carbon-enhanced metal-poor stars as a function of declining metallicity[45], and as a function of distance from the Galactic plane[46,47].

Much remains to be learned as the database of low-metallicity stars, in particular those that are found in distant *in situ* samples, or from those nearby stars with available proper motions that indicate membership of the outer-halo population. We look forward to the next dramatic increase in the numbers of very metal-poor stars that will come from the ongoing stellar samples from SDSS, in particular from SEGUE, the Sloan Extension for Galactic Understanding and Exploration.

**Supplementary Information** is linked to the online version of the paper at www.nature.com/nature.

**Acknowledgements** We thank C. Allende Preito, E. Bell, W. Brown, A. Frebel, B. Gibson, H. Morrison, C. Thom, J. Tumlinson, and B. Yanny for comments received on previous versions of this Article. D.C. acknowledges partial support for travel and living expenses from JINA, the Joint Institute for Nuclear Astrophysics, while in residence at Michigan State University. Funding for the SDSS and SDSS-II has been provided by the Alfred P. Sloan Foundation, the Participating Institutions, the National Science Foundation, the US Department of Energy, the National Aeronautics and Space Administration, the Japanese Monbukagakusho, the Max Planck Society, and the Higher Education Funding Council for England. The SDSS Web Site is http://www.sdss.org.

**Author Information** Reprints and permissions information is available at www.nature.com/reprints.

Correspondence and requests for materials should be addressed to D.C. (carollo@mso.anu.edu.au).

## Table 1 Studies claiming a retrograde outer halo

| Sample and selection criteria | N | Additional Restrictions | $<V_\Phi>$ (km s$^{-1}$) | Method | Source |
|---|---|---|---|---|---|
| Globular clusters (non-kinematic) | 19 | 'Young halo' | −64 ± 74 | F & W | Ref. 2 |
| Globular clusters (non-kinematic) | 20 | 'Young halo' | −42 ± 80 | F & W | Ref. 10 |
| RR Lyraes (non-kinematic) | 26 | \|Z\| < 8 kpc | −95 ± 29 | FSM | Ref. 9 |
| Field subdwarfs (kinematic) | 30 | $Z_{max}$ > 5 kpc | −45 ± 22 | FSM | Ref. 7 |
| | | Bias Corrected | +24 ± 13 | | Ref. 16 |
| Field horizontal-branch stars (non-kinematic) | 90 | [Fe/H] < −1.6 \|Z\| > 4 kpc | −93 ± 36 | F & W | Ref. 8 |
| Field subdwarfs (kinematic) | 101 | V < −100 km s$^{-1}$ [Fe/H] < −1.8 | −32 ± 10 | FSM | Ref. 13 |
| Field F,G,K dwarfs (non-kinematic) | 250 | \|Z\| > 5 kpc | −55 ± 16 | FSM | Ref. 6 |
| Field F,G turnoff (non-kinematic) | | $Z_{max}$ > 5 kpc | | FSM | This work |
| | 2228 | [Fe/H] < −1.0 | −11 ± 2 | | |
| | 200 | [Fe/H] < −2.2 | −41 ± 11 | | |
| | | $Z_{max}$ > 10 kpc | | | |
| | 771 | [Fe/H] < −1.0 | −38 ± 5 | | |
| | 94 | [Fe/H] < −2.2 | −71 ± 17 | | |
| | | $Z_{max}$ > 15 kpc | | | |
| | 371 | [Fe/H] < −1.0 | −56 ± 8 | | |
| | 54 | [Fe/H] < −2.2 | −71 ± 25 | | |

**Table Caption |** Previous and current determinations of the mean rotational velocity, $<V_\Phi>$, and the error in the mean $(\sigma / \sqrt{N}$ , where σ is the standard deviation and N is the number of stars) for samples in which a counter-rotating halo has been claimed, ordered by sample size. The samples listed in the first column are classified as to whether they were chosen on the basis of high proper motions (kinematic) or not (non-kinematic). Restrictions placed on each sample by the authors are listed in the third column (see original papers for details). The method of analysis used for each determination is listed: F & W, estimate based on the technique of Frenk & White[48] , which considers distances and radial velocities alone, under the assumption of a cylindrically symmetric Galaxy; FSM, estimate based on consideration of the full space motions, which requires the use of proper motions, as well as distances and radial velocities. The samples analyzed with the F & W approach are either not statistically different from zero (refs 2, 10), or are only marginally so (2.6σ; ref. 8). Previous samples based on analysis of the full space motions vary from statistically insignificant (ref. 7), to just over 3σ significance (refs. 6, 9, 13), due to the small numbers of stars considered. Note that after application of (uncertain) corrections for kinematic bias (ref. 16), the retrograde result reported in ref. 7 disappears entirely. One can assume that a similar outcome might apply to the ref. 13 determination. The samples of ref. 6 and ref. 9 are both selected over a restricted region of sky (towards the North Galactic Pole) and are therefore subject to possible contamination by individual stellar streams. All but two subsamples of the SDSS calibration star sample have retrograde signals that are significant at more than the 4σ level. The subsample at $Z_{max} > 5$ kpc and $[Fe/H] < -1$ is likely to include significant contamination from inner-halo stars.

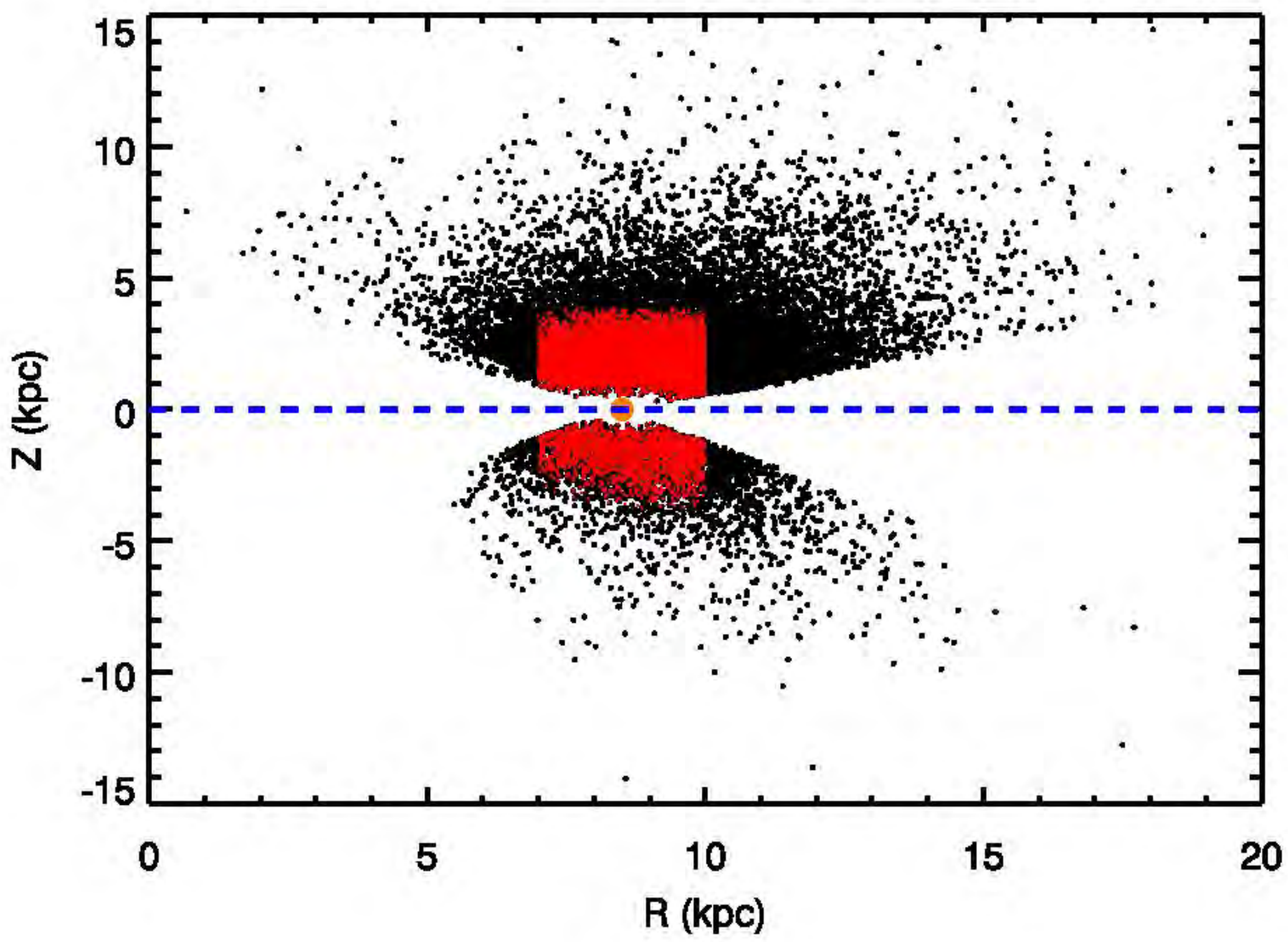

**Figure 1 | The spatial distribution of the stars analyzed in the present sample.** The distribution of the full sample of 20,236 unique SDSS (Data Release 5[12]) spectrophotometric and telluric calibration stars in the Z-R plane is shown, where Z is the derived distance from the Galactic plane in the vertical direction and R is the derived distance from the centre of the Galaxy projected onto this plane. The dashed blue line represents the Galactic plane, while the filled orange dot is the position of the Sun, at Z = 0 kpc and R = 8.5 kpc. The 'wedge shape' of the selection area is the result of limits of the SDSS footprint in Galactic latitude. The red points indicate the 10,123 stars that satisfy our criteria for a local sample of stars, having 7 kpc < R < 10 kpc, with distance estimates from the Sun d < 4 kpc, and with viable measurements of stellar parameters and proper motions.

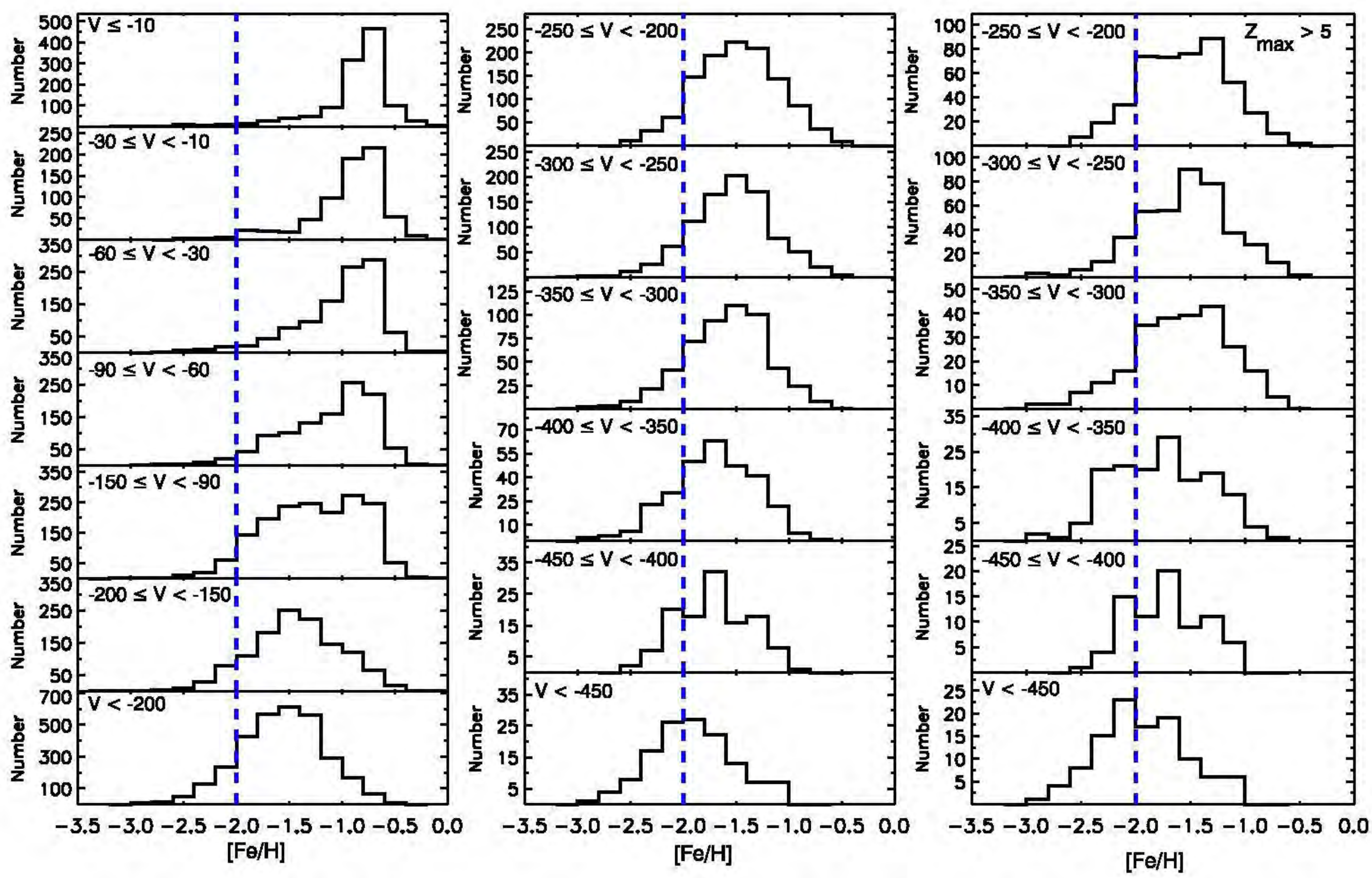

**Figure 2 | The distribution of [Fe/H] for various cuts in the V velocity, the component of orbital motion measured with respect to the Local Standard of Rest.** The Local Standard of Rest is a frame in which the mean space motions of the stars in the Solar neighbourhood average to zero. A blue dashed line at [Fe/H] = −2.0 is added for reference in all three columns. In the left-hand column, the full data set is considered. The stars with modestly negative V velocities in the upper three panels are dominated by stars from the thick-disk (and metal-weak thick-disk) populations, with a peak metallicity around [Fe/H] ≈ −0.7. A transition to dominance by inner- and outer-halo population stars becomes evident for V < −90 km $s^{-1}$; in the bottom panel of this column, the distribution of [Fe/H] appears similar to what in the past was considered ‘the halo’, but we argue results from a superposition of contributions from both inner- and outer-halo populations. In the middle column, the large numbers of stars with V < −200 km $s^{-1}$ are broken into smaller ranges in V velocity. As V becomes increasingly retrograde (V < −220 km $s^{-1}$), the metallicity distribution shifts to include ever larger numbers of stars with [Fe/H] < −2.0, and relatively

fewer stars with [Fe/H] ≈ −1.6. The same V velocity cuts are applied in the right-hand column, but only for stars with $Z_{max} > 5$ kpc, in order to decrease the contribution from inner-halo population stars. Although fewer stars are included, the increasing dominance of stars with [Fe/H] < −2.0 is even more apparent. We associate the stars with the most extreme retrograde orbits (and those that reach far above the Galactic plane in their orbits) with the outer-halo population.

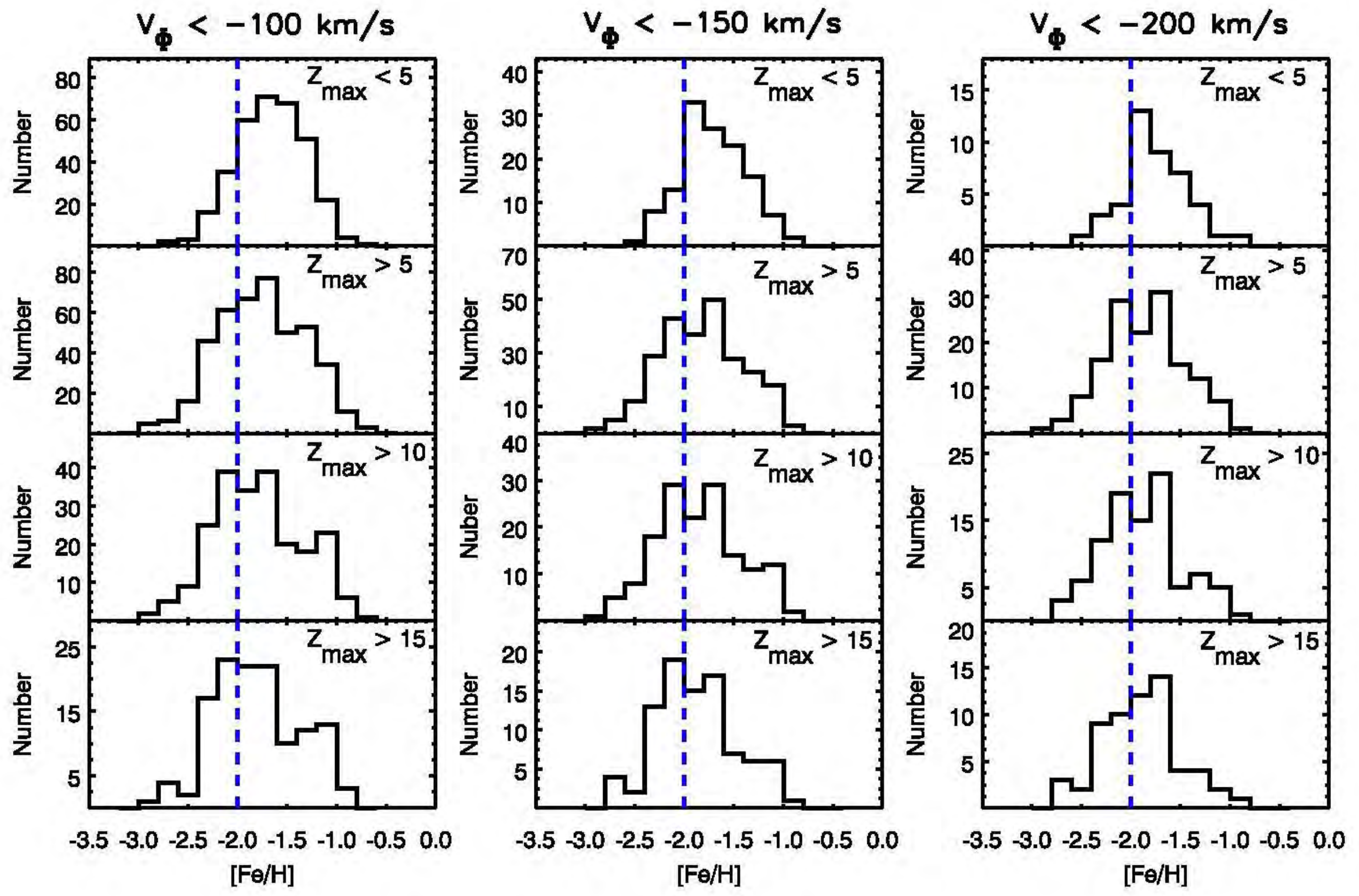

**Figure 3 | The distribution of [Fe/H] for the stars in our sample on highly retrograde orbits.** Stars from the disk populations, which possess prograde orbits, cannot be present in this plot. The panels show various cuts in $V_\Phi$, the rotational velocity with respect to the Galactic Centre in a cylindrical coordinate system, and for different ranges of $Z_{max}$ (in kpc). A blue dashed line at [Fe/H] = −2.0 is added for reference. The left-hand column applies for stars with $V_\Phi < -100$ km s$^{-1}$; the clearly skewed distribution of [Fe/H] exhibits an increased contribution from lower metallicity stars as one progresses from the low ($Z_{max} < 5$ kpc) to the high ($Z_{max} > 15$ kpc) subsamples. Simultaneously, the predominance of stars from the inner-halo population, with peak metallicity at [Fe/H] ≈ −1.6, decreases in relative strength, and shifts to lower [Fe/H]. Similar behaviours are seen in the middle and right-hand columns, which correspond to cuts on $V_\Phi < -150$ km s$^{-1}$ and −200 km s$^{-1}$, respectively. A Kolmogorov-Smirnoff test of the null hypothesis that the MDFs of stars shown in the lower panels for the individual cuts on $V_\Phi$ could be drawn from the MDFs of the same parent population as those shown in the upper panels, against an alternative that the stars are drawn from more metal-

poor parent MDFs, is rejected at high levels of statistical significance. For $V_{\Phi} < -100$ km $s^{-1}$, one-sided probabilities less than 0.0001 are obtained for the cuts on $Z_{max} > 5$, 10, and 15 kpc, respectively. For $V_{\Phi} < -150$ km $s^{-1}$, one-sided probabilities of 0.0004, 0.0001, and 0.0003 are obtained, for $Z_{max} > 5$, 10, and 15 kpc, respectively. For $V_{\Phi} < -200$ km $s^{-1}$, one-sided probabilities of 0.014, 0.010, and 0.033 are obtained, for $Z_{max} > 5$, 10, and 15 kpc, respectively.

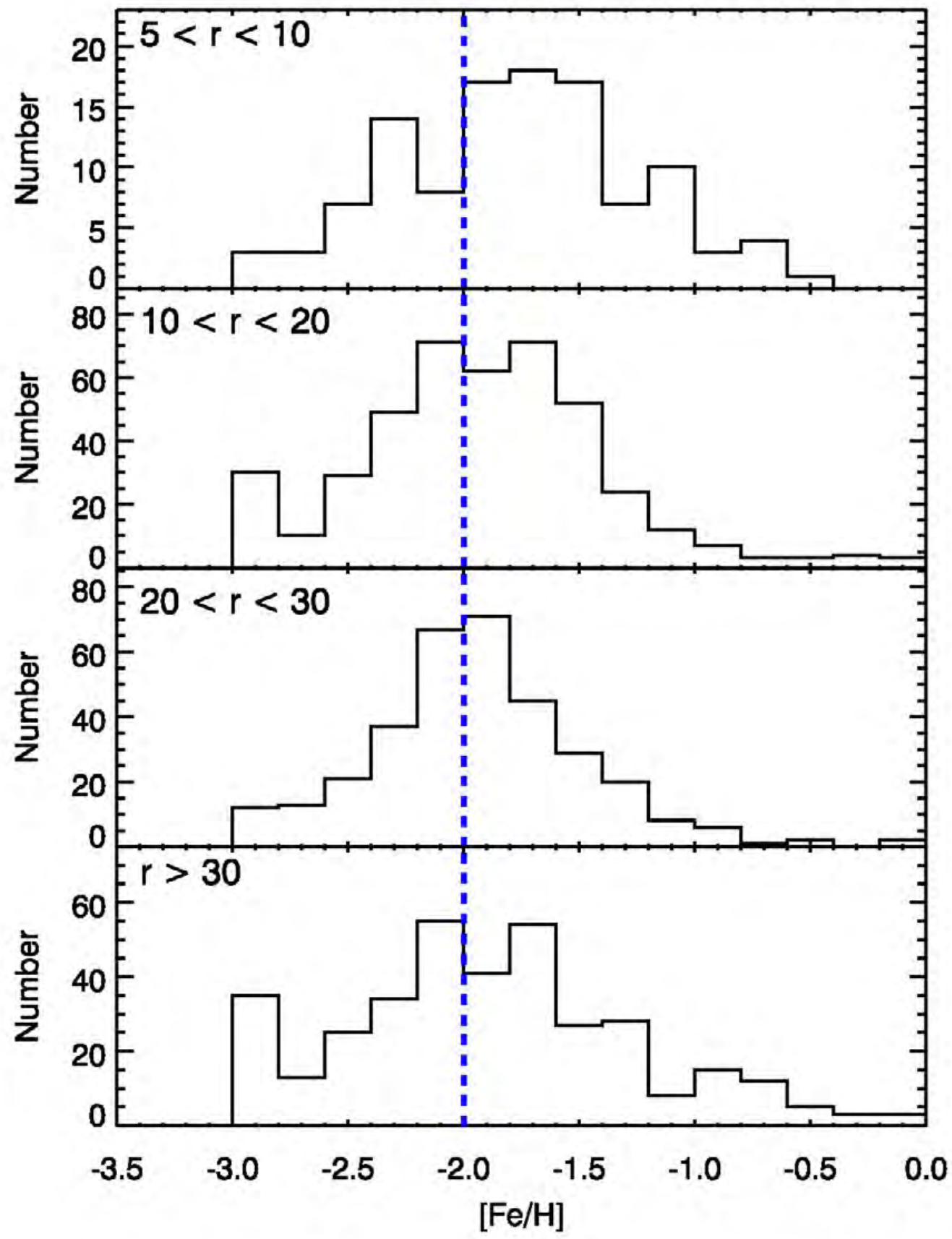

**Figure 4 | A sample of blue horizontal-branch stars exploring much larger distances from the Galactic Centre than the SDSS calibration stars.** The distribution of [Fe/H] is shown for various cuts on the distance from the Galactic Centre, r, in kpc. The nature of the MDF appears to shift from the upper two panels, which exhibit the character of a mixture of inner- and outer-halo populations, over to a unimodal distribution in the third panel, centred on [Fe/H] ≈ −2.0. The most distant blue horizontal-branch (BHB) stars in the lowest panel also exhibit the appearance of a mixture of the two populations, possibly due to the inclusion of inner-halo stars on highly eccentric orbits that take them far from the Galactic Centre. The peak around [Fe/H] = −3.0 seen in several of the panels is an artefact arising from the limit of the metallicity grid that is used for abundance determinations of the BHB stars. A Kolmogorov-Smirnoff test of the null hypothesis that the MDFs of stars shown in the lower

panels for the individual cuts on Galactocentric distance r could be drawn from the same parent population as the stars shown in the first panel, against an alternative that the stars are drawn from more metal-poor parent MDFs, is rejected at high levels of statistical significance (one-sided probabilities of 0.0262, 0.0005, and 0.0243, respectively, for the three higher cuts on Galactocentric distance). The fraction of stars with metallicities $[Fe/H] < -2.0$ (primarily outer-halo stars) grows from 31% for stars with $5 < r < 10$ kpc to 46% for stars at larger Galactocentric distances.

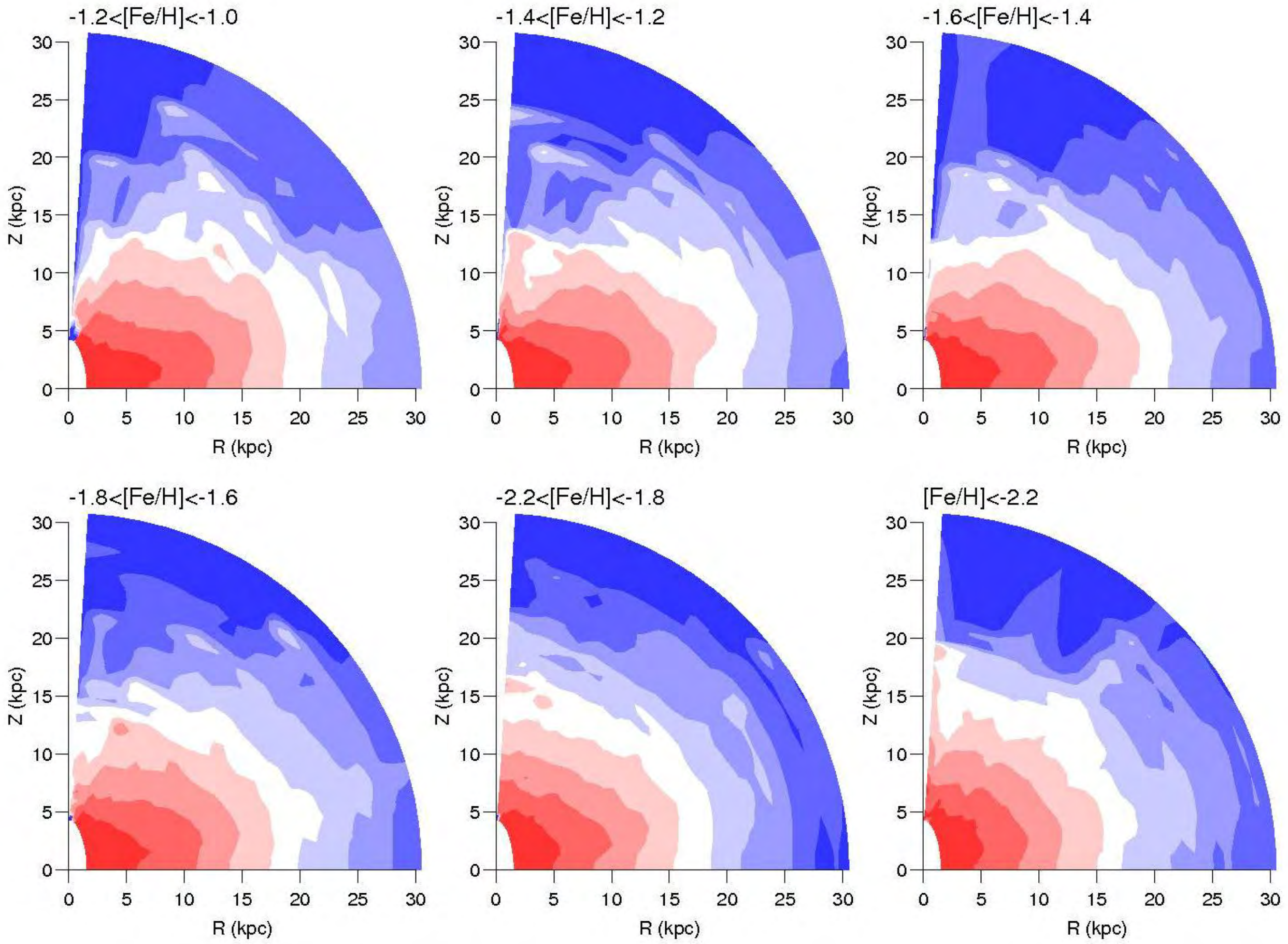

**Figure 5 | Equidensity contours of the reconstructed global density distributions for stars in our sample with various metallicities.** The global density distributions are constructed from the sum of the probability density of an orbit at each location in the Z-R plane, with a weighting factor being inversely proportional to the corresponding density at the currently observed position of the star[19,20]. High-density regions are indicated by redder colours, while low-density regions are indicated by bluer colours (a linear density scale is employed). Within each metallicity cut the apparent flattening of the inner regions slowly goes over to a more spherical shape with increasing distance. As one progresses from the more metal-rich ([Fe/H] $\approx -1.0$) to the most metal-poor ([Fe/H] $< -2.2$) subsets of these data, the overall nature of the equidensity contours also changes from highly flattened (axial ratios

of $\sim 0.6$), to more spherical (axial ratio of $\sim 0.9$). This suggests that the inner- and outer-halo components are broadly overlapping in space and in metallicity – the inner-halo population is characterized as a flattened density distribution that dominates locally for stars with [Fe/H] $> -2$, whereas the outer-halo population is nearly spherical, and dominates at larger distances and locally for stars with [Fe/H] $< -2$.

# SUPPLEMENTARY INFORMATION

## Selection of the SDSS Calibration Stars

During the course of the original SDSS it was necessary to obtain medium-resolution ($R = \lambda/\Delta\lambda = 2000$) spectroscopy of a small number (16) of calibration stars per spectroscopic plug-plate (when fully populated with fibres, one plug-plate obtains 640 spectra simultaneously), chosen for two primary reasons. The first set of these objects, the spectrophotometric calibration stars, are stars that are selected to approximately remove the distortions of the observed flux of stars and galaxies arising from the wavelength response of the ARC 2.5m telescope and the SDSS spectrographs, as well as the distortions imposed on the observed spectra by the Earth's atmosphere. For this purpose, there is an advantage in choosing stars that satisfy three conditions: (1) their spectra are relatively clean of absorption lines arising from metallic species in the stars, (2) they are sufficiently bright, so that high signal-to-noise (S/N > 20/1) spectra can be obtained, and (3) they exhibit spectra that can be reasonably well-modelled by synthetic spectra on a known flux scale for a known class of stars, in this case, stars close to the halo main-sequence turnoff with metallicities near [Fe/H] = −2.0. These goals lead to a selection of stars with apparent magnitudes in the range $15.5 < g_0 < 17.0$, and satisfying the colour ranges $0.6 < (u\text{-}g)_0 < 1.2$ ; $0 < (g\text{-}r)_0 < 0.6$. The subscript 0 in the magnitudes and colours indicates that they are corrected for the effects of interstellar absorption and reddening, following standard procedures[49]. The colour range is intentionally kept rather large, so that possible spectrophotometric calibration stars may be found even in cases where the density of stars is low, such as at the Galactic poles. Within these constraints, efforts are made to select stars that are as blue as possible, as these are the most likely to be consistent with [Fe/H] = −2.0. The second set of calibration stars, the telluric calibration stars, are used in order to calibrate and remove from SDSS spectra the presence of night sky emission and absorption features. The telluric calibration stars cover the same colour ranges as the spectrophotometric calibration stars, but at fainter apparent magnitudes, in the range $17.0 < g_0 < 18.5$. The requirement for favouring the bluer stars is also relaxed.

The S/N of the resulting spectra is lower than for the spectrophotometric stars owing to their fainter apparent magnitudes. Generally, the achieved range is $20/1 < S/N < 30/1$ for these stars. The full sample comprises 20,236 unique stars with acceptable derived atmospheric parameters.

Although there is clearly a bias towards the identification of metal-poor stars arising from the colour selections (in particular for the spectrophotometric calibration stars), it should be kept in mind that it is not possible to discriminate the *lowest* metallicity stars, e.g., those with [Fe/H] $< -2.5$, from those with [Fe/H] $\sim -2.0$, since the effect of declining metallicity on broadband stellar colours is minimal in this regime. Hence, we expect that the distribution of metallicities for the calibration stars should reflect the true shape of the low-metallicity tail of the MDF for stars outside the disk populations. The same does not apply for stars with [Fe/H] $> -2.0$; the observed distributions of metallicity for such stars will mis-represent the true MDFs at these abundances. It is important to note that no bias on the kinematics, e.g., by making use of measured proper motions (the angular displacement of a given star over time, in the direction perpendicular to the line of sight), is introduced in the selection of the calibration objects. This choice enables our kinematic studies, without the need to apply explicit corrections or modeling of any selection bias for the motions of the stars in the sample.

Once estimates of the atmospheric parameters for each of the calibration stars are obtained (as described below), one can use the derived surface gravity of each star to infer whether it is a likely dwarf, subgiant, or giant. Photometric estimates of the distance to each star (accurate to ~ 10-20%) are then obtained by comparison of its observed apparent magnitude (corrected for interstellar absorption) with its expected absolute magnitude based on calibrated open cluster and globular cluster sequences[50].

## Stellar Atmospheric Parameter Analysis

The SEGUE Stellar Parameter Pipeline (SSPP)[51,52,53,54,55] processes the wavelength- and flux-calibrated spectra generated by the standard SDSS spectroscopic reduction pipeline[56], obtains equivalent widths and/or line indices for about 80 atomic or molecular absorption lines, and estimates the effective temperature, $T_{eff}$, surface gravity, log g, and [Fe/H] for a given star through the application of a number of approaches. The current techniques employed by the SSPP include chi-square minimization with respect to synthetic spectral templates[51,53,57], neural network analysis[58], auto-correlation analysis[59], and a variety of line-index calculations based on previous calibrations with respect to known standard stars[59]. The SSPP employs five primary methods for the estimation of $T_{eff}$, eight for the estimation of log g, and nine for the estimation of [Fe/H]. The use of multiple methods allows for empirical determinations of the internal errors for each parameter, based on the range of reported values from each method – typical *internal* errors for stars in the temperature range that applies to the calibration stars are $\sigma$ ($T_{eff}$) ~ 100 K to 125 K, $\sigma$ (log g) ~ 0.25 dex, and $\sigma$ ([Fe/H]) ~ 0.20 dex. Over the past two years, large-aperture telescopes have been used to obtain high-resolution spectroscopy for over 150 of the brighter ($14 < g_0 < 16.0$) SDSS stars. Analysis of these data indicates that the *external* errors on the determinations of atmospheric parameters generated by the SSPP are in fact of similar magnitude[53,54,55,60].

## Derivation of Kinematic Parameters

In order to obtain the best available estimates of the kinematics and orbital parameters for the stars in our sample as a function [Fe/H], we consider only those stars satisfying several cuts: (1) a selection in the effective temperature range 5000 K < $T_{eff}$ < 6800 K, over which the SSPP is expected to provide the highest accuracy in both the atmospheric parameters and chemical compositions, reducing the number of stars to 19,592, (2) a selection of the stars in the sample with distances d < 4 kpc from the Sun, in order to restrict the kinematical and orbital analyses to a local volume (where the assumptions going into their calculation are best satisfied), reducing the number of stars to 14,175. This selection also mitigates against the increase in the errors in the derived transverse velocities, which scale with distance from the Sun (e.g., for a typical star in our sample, at d = 1 kpc, errors of 3.5 mas/yr in proper motions and 15% in distance result in errors in the derived transverse velocities of 22 km/s; these rise to 37 km/s and 46 km/s for stars at d = 2 kpc and 4 kpc, respectively) and (3) a selection on R, the present Galactocentric distance of a star projected onto the Galactic plane, to be in the range $7 < R < 10$ kpc. After these restrictions are applied, the number of stars in the remaining sample is 10,123.

Radial velocities for stars in our sample are derived from matches to an external library of high-resolution spectral templates with accurately known velocities[61], degraded in resolution to that of the SDSS spectra. The typical accuracies of the resulting radial velocities are on the order of 3-20 km/s, depending on the S/N of the spectra, with zero-point errors no more than 3 km/s, based on a comparison of the subset of stars in our sample with radial velocities obtained from the high-resolution spectra taken for testing and validation of the SSPP[55].

The proper motions (taken from the re-calibrated USNO-B catalogue, with typical accuracy around 3-4 mas/yr[62]), in combination with the distance estimates and radial velocities, provide the information required to calculate the full space motions. The components of the space motions are represented as the (U,V,W) velocities of the stars with respect to the Local Standard of Rest (LSR; defined as a frame in which the mean space motions of the stars in the Solar neighbourhood average to zero). The velocity component U is taken to be positive in the direction toward the Galactic anti-centre, the V component is positive in the direction toward Galactic rotation, and the W component is positive toward the north Galactic pole. Corrections for the motion of the Sun with respect to the LSR are applied during the course of the calculation of the full space motions; here we adopt the values $(U_{\odot}, V_{\odot}, W_{\odot}) = (-9, 12, 7)$ km/s[63].

For our analysis it is also convenient to obtain the rotational component of a star's motion about the Galactic centre in a cylindrical frame; this is denoted as $V_{\Phi}$ , and is calculated assuming that the LSR is on a circular orbit with a value of 220 km/s[64]. The orbital parameters of the stars, such as the perigalactic distance (the closest approach of a stellar orbit to the Galactic centre, $r_{peri}$), the apogalactic distance (the farthest extent of a stellar orbit from the Galactic centre, $r_{apo}$), the orbital eccentricity, and $Z_{max}$ (the maximum distance of stellar orbits above or below the Galactic plane) are derived by adopting an analytic Stäckel-type gravitational potential (which consists of a flattened, oblate disk and a nearly spherical massive halo) and integrating their orbital paths based on the starting point obtained from the observations[65].

**Kinematics of the SDSS Calibration Stars**

Supplemental Figure 1 shows the three velocity components (U,V,W) as a function of the derived metallicity, [Fe/H]. One can clearly distinguish the presence of the low velocity dispersion thick-disk and metal-weak thick-disk populations[65], as well as a second group of stars that exhibits a broad dispersion in both their velocity components and metallicities; these stars comprise the overlapping inner- and outer-halo populations.

Supplemental Table 1 summarizes the derived kinematics of our sample as a whole, as well as for different cuts in the present distance of stars above or below the Galactic plane. The errors on the derived <U,V,W> and on their dispersions, σ (U,V,W), are the most precise estimates obtained to date, owing to the very large numbers of stars in our sample.

Supplemental Table 2 provides an alternative set of estimated net rotation for the subsamples considered in Supplemental Table 1. These estimates are derived on the basis of radial velocities and distances alone (and through adoption of an axi-symmetric model for the distribution of stars in the Galaxy)[66], so that errors in the measured proper motions cannot propagate into the inferred values of rotation (the net rotation about the Galactic centre obtained from such an analysis is conventionally referred to as $V_{rot}$). As is clear from inspection of this table (Supplemental Fig. 2 provides a graphical representation of these results), the values of rotation obtained from the analysis of the full space motions are generally in good agreement with the reported $V_{rot}$, although the derived errors in $V_{rot}$ become quite large when the number of stars is small. This provides a demonstration that any systematic errors in the derived proper motions used in our analysis must be rather small, and furthermore, that they do not scale with distance.

Toomre (as attributed by ref. 67) introduced a convenient method for visualizing the distribution of orbital energies for stars observed in the local neighbourhood, by plotting the parameter $[U^2 + W^2]^{1/2}$ as a function of the V velocity component. Sets of Toomre diagrams for the stars in our sample are shown in Supplemental Fig. 3, for various cuts on $Z_{max}$ and [Fe/H]. The different cuts on these quantities demonstrates that there exists an imbalance in the distribution of orbital energies for stars with $Z_{max} < 5$ kpc, and for stars in the intermediate interval 5 kpc $< Z_{max} <$ 10 kpc, in the sense that the stars on prograde orbits populate lower-energy orbits than do the stars with retrograde orbits. Only when one considers the lower panels of Supplemental Fig. 3, which correspond to stars with $Z_{max} > 10$ kpc, does it appear that there the distribution of orbital energies between the prograde and retrograde stars is roughly similar at all metallicities. This behaviour is a result of the highly overlapped inner- and outer-halo populations in the Solar neighbourhood.

The use of metallicity and kinematics to constrain models of the formation of the Galaxy began over four decades ago[68]. The relatively clear correlation between [Fe/H] and eccentricity in Fig. 4 of ref. 68 was interpreted as providing strong evidence for a rapid, monolithic collapse of the Galaxy, with stars forming on a similar timescale as that of the collapse, on the order of no more than 1 Gyr. In this model, the motions of lower metallicity stars reflected earlier gas dynamical evolution of the proto-Galactic cloud, inside of which star formation, and thus chemical evolution, proceeded progressively. The adoption of orbital eccentricity as a dynamical indicator for stars is advantageous, as this is nearly an adiabatic invariant under a slow change of a gravitational potential[68]. It has since been shown that the appearance of the [Fe/H] vs. eccentricity diagram was strongly influenced by the kinematic selection criteria used by these authors. The equivalent diagram for a larger *non-kinematically* selected sample of stars (see Fig. 6a of ref. 65) clearly demonstrated that stars are found in the Galaxy over the entire range of observed [Fe/H] and orbital eccentricity[65]. Supplemental Figure 4 shows this diagram for the sample of SDSS calibration stars considered herein; the number of stars shown represents an order of magnitude increase in sample size as compared to all previous samples. Although one can clearly discern concentrations of stars with higher metallicity and lower eccentricity, as well as stars with lower metallicity and higher eccentricity, there exist many stars for which these two quantities appear uncorrelated. A monolithic collapse model would not predict such behaviour.

We now consider how the distribution of [Fe/H] changes for stars as a function of their orbital eccentricities. The left column of Supplemental Fig. 5 applies for stars at all $Z_{max}$, while the right column applies for stars with $Z_{max} > 5$ kpc. Very different behaviours in the distribution of [Fe/H] are found when comparing the left-hand and right-hand columns in various ranges of orbital eccentricity. The change in the distribution of [Fe/H] in the left-hand column could be fully explained by considering the transition from stars with disk-like metallicities (and low-eccentricity orbits) to those with a single-component halo distribution of metallicities and high-eccentricity orbits. However, the right-hand column, which applies to stars with $Z_{max} > 5$ kpc, exhibits a behaviour that has not been seen previously. The upper panel of this column shows a clear bimodal distribution of [Fe/H], due to the presence of outer-halo stars with low-eccentricity orbits. The middle panels exhibit a broad distribution of [Fe/H], which we interpret as an overlap of contributions from the inner- and outer-halo populations, and the lower panels show the dominance of the contribution from inner-halo stars. This would not be expected if the halo were a single stellar population.

Our analysis also indicates that the inner- and outer-halo populations exhibit different orbital characteristics. Supplemental Figure 6 shows the distribution of [Fe/H] for stars as a function of the derived $r_{peri}$ and $r_{apo}$ distances for various cuts in the value of $V_\Phi$ . It is clear that the majority of inner-halo stellar orbits penetrate close to the Galactic centre (within about 3 kpc), while the outer-halo stellar orbits (those stars on highly retrograde orbits) rarely penetrate this close. The majority of inner-halo stars also exhibit orbits that reach no farther than about 15-20 kpc from the Galactic centre, whereas the outer-halo stars cover the range in apogalactic distance between 10 and 80 kpc fairly uniformly. It should be kept in mind that in order to appear in our sample, stars must be on orbits that pass through the Solar neighbourhood. Thus, there exists the possibility that some degree of ``Keplerian censorship'' applies (as pointed out by an anonymous referee). The full extent of this orbital censorship, and its influence on the appearance of Supplemental Fig. 6, requires additional modeling. Regardless of the degree of orbital censorship that might exist, the shift in the mean metallicity from the higher metallicity inner-halo to the lower metallicity outer-halo population is clear.

**Supplementary References**

**Supplemental Table 1   Full space motions for stars in the local volume**

| Sample | [Fe/H] Range | N | <U> (km/s) | <V> (km/s) | <W> (km/s) | σ(U) (km/s) | σ (V) (km/s) | σ (W) (km/s) | $<V_\Phi>$ (km/s) |
|---|---|---|---|---|---|---|---|---|---|
| All Stars | -0.40 to -0.80 | 1853 | -24 ± 1 | -52 ± 1 | -2 ± 1 | 58 ± 1 | 60 ± 1 | 40 ± 1 | 170 ± 1 |
| | -0.80 to -1.00 | 1616 | -23 ± 2 | -78 ± 2 | -3 ± 1 | 81 ± 1 | 77 ± 1 | 54 ± 1 | 143 ± 2 |
| | -1.00 to -1.40 | 2496 | -19 ± 3 | -153 ± 2 | -2 ± 1 | 132 ± 2 | 102 ± 1 | 73 ± 1 | 68 ± 2 |
| | -1.40 to -1.80 | 2491 | -14 ± 3 | -199 ± 2 | 0 ± 2 | 152 ± 2 | 107 ± 2 | 81 ± 1 | 22 ± 2 |
| | -1.80 to -2.20 | 1260 | -21 ± 4 | -224 ± 3 | 1 ± 3 | 158 ± 3 | 119 ± 2 | 103 ± 2 | -2 ± 3 |
| | < -2.20 | 349 | -25 ± 9 | -241 ± 8 | 8 ± 6 | 169 ± 6 | 159 ± 6 | 120 ± 5 | -19 ± 8 |
| \|Z\| < 1 kpc | -0.40 to -0.80 | 117 | -4 ± 4 | -14 ± 3 | 8 ± 3 | 46 ± 3 | 35 ± 2 | 32 ± 2 | 206 ± 3 |
| | -0.80 to -1.00 | 108 | -1 ± 7 | -48 ± 7 | -4 ± 4 | 70 ± 5 | 70 ± 5 | 44 ± 3 | 172 ± 7 |
| 1 < \|Z\| < 2 kpc | -1.00 to -1.40 | 1173 | -10 ± 4 | -136 ± 3 | -5 ± 2 | 123 ± 3 | 99 ± 2 | 63 ± 1 | 85 ± 3 |
| | -1.40 to -1.80 | 1211 | -8 ± 4 | -193 ± 3 | -5 ± 2 | 144 ± 3 | 106 ± 2 | 79 ± 1 | 28 ± 3 |
| | -1.80 to -2.20 | 502 | -14 ± 7 | -216 ± 5 | -2 ± 5 | 147 ± 5 | 114 ± 4 | 95 ± 2 | 5 ± 5 |
| | < -2.20 | 115 | 1 ± 14 | -232 ± 14 | 11 ± 10 | 153 ± 10 | 147 ± 10 | 124 ± 5 | -13 ± 14 |
| 2 < \|Z\| < 3 kpc | -1.00 to -1.40 | 971 | -32 ± 4 | -170 ± 3 | -1 ± 3 | 139 ± 3 | 103 ± 2 | 78 ± 2 | 51 ± 3 |
| | -1.40 to -1.80 | 944 | -17 ± 5 | -205 ± 3 | 3 ± 3 | 161 ± 4 | 107 ± 2 | 84 ± 2 | 17 ± 3 |
| | -1.80 to -2.20 | 561 | -31 ± 7 | -226 ± 5 | 2 ± 4 | 166 ± 5 | 117 ± 4 | 99 ± 3 | -4 ± 5 |
| | < -2.20 | 182 | -36 ± 14 | -236 ± 12 | 12 ± 9 | 187 ± 10 | 159 ± 8 | 120 ± 6 | -12 ± 11 |
| 3 < \|Z\| < 4 kpc | -1.00 to -1.40 | 217 | -34 ± 10 | -191 ± 7 | 5 ± 6 | 154 ± 7 | 104 ± 5 | 82 ± 4 | 28 ± 7 |
| | -1.40 to -1.80 | 208 | -45 ± 11 | -218 ± 8 | 4 ± 7 | 159 ± 8 | 121 ± 6 | 95 ± 5 | 2 ± 8 |
| | -1.80 to -2.20 | 141 | -9 ± 15 | -254 ± 12 | 8 ± 10 | 177 ± 11 | 138 ± 8 | 122 ± 7 | -34 ± 12 |
| | < -2.20 | 43 | -60 ± 18 | -285 ± 28 | -9 ± 21 | 121 ± 13 | 184 ± 20 | 139 ± 15 | -63 ± 28 |

**Supplemental Table 1 Caption |** Kinematic quantities for stars in the local volume based on full space motions. The estimates of <U,V,W> and σ (U,V,W) are derived making use of the combination of measured proper motions and radial velocities, as well as estimated distances. The errors listed are the error in the mean ($\sigma/\sqrt{N}$). For stars with |Z| < 1 kpc one can clearly recognize thick-disk-like rotation and velocity dispersions, while at higher |Z| the rotation and dispersions result from the overlap of the inner- and outer-halo populations.

## Supplemental Table 2 A Frenk & White Analysis of Rotational Velocities

| Sample | [Fe/H Range | N | <$V_\Phi$> (km/s) | $V_{rot}$ (km/s) |
|---|---|---|---|---|
| All Stars | -0.40 to -0.80 | 1853 | 170 ± 1 | 156 ± 3 |
| | -0.80 to -1.00 | 1616 | 143 ± 2 | 131 ± 3 |
| | -1.00 to -1.40 | 2496 | 68 ± 2 | 57 ± 4 |
| | -1.40 to -1.80 | 2491 | 22 ± 2 | 11 ± 4 |
| | -1.80 to -2.20 | 1260 | -2 ± 3 | 12 ± 6 |
| | < -2.20 | 349 | -19 ± 8 | -10 ± 13 |
| \|Z\| < 1 kpc | -0.40 to -0.80 | 117 | 206 ± 3 | 192 ± 7 |
| | -0.80 to -1.00 | 108 | 172 ± 7 | 154 ± 12 |
| 1 < \|Z\| < 2 kpc | -1.00 to -1.40 | 1173 | 85 ± 3 | 73 ± 5 |
| | -1.40 to -1.80 | 1211 | 28 ± 3 | 12 ± 6 |
| | -1.80 to -2.20 | 502 | 5 ± 5 | 12 ± 9 |
| | < -2.20 | 115 | -13 ± 14 | -12 ± 20 |
| 2 < \|Z\| < 3 kpc | -1.00 to -1.40 | 971 | 51 ± 3 | 37 ± 7 |
| | -1.40 to -1.80 | 944 | 17 ± 3 | 8 ± 7 |
| | -1.80 to -2.20 | 561 | -4 ± 5 | 16 ± 10 |
| | < -2.20 | 182 | -12 ± 11 | 0 ± 18 |
| 3 < \|Z\| < 4 kpc | -1.00 to -1.40 | 217 | 28 ± 7 | 22 ± 17 |
| | -1.40 to -1.80 | 208 | 2 ± 8 | -1 ± 22 |
| | -1.80 to -2.20 | 141 | -34 ± 12 | -17 ± 28 |
| | < -2.20 | 43 | -63 ± 28 | -106 ± 56 |

**Supplemental Table 2 Caption |** Rotation estimates for stars in the local volume. The Frenk & White[66] analysis employed for the calculation of the net rotation, $V_{rot}$, makes use of the measured radial velocities and distance estimates, so is not subject to possible systematic and random errors arising from use of the measured proper motions. An axi-symmetric Galaxy is also assumed. For comparison, the value of <$V_\Phi$> from Supplemental Table 1, based on analysis of the full space motions, is listed for the same subsamples of stars. The errors listed are the error in the mean ($\sigma / \sqrt{N}$). Note that the estimates of net rotation are, in most cases, commensurate within the quoted errors.

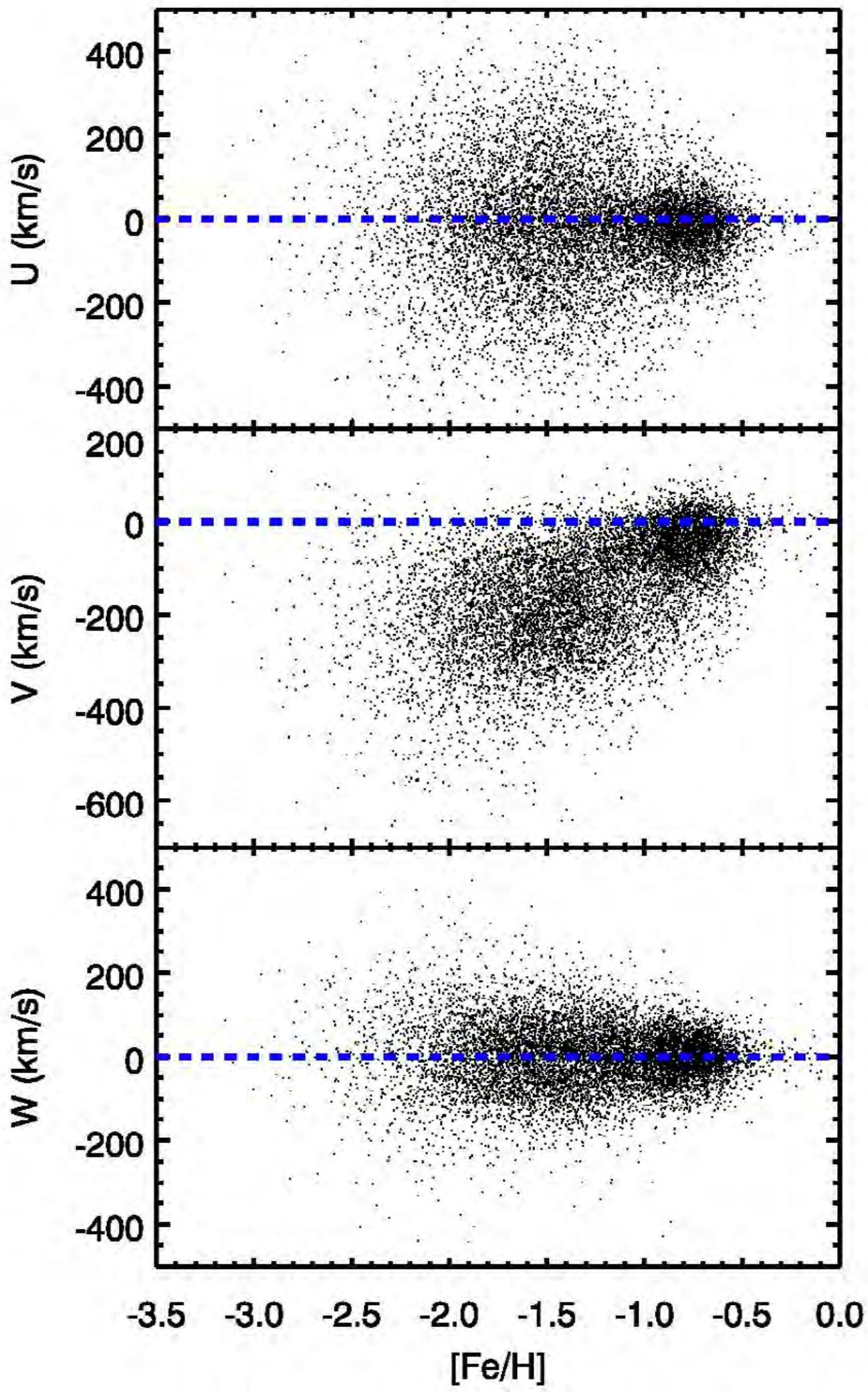

**Supplemental Figure 1 |** The U, V, and W components of the space motions of stars in our sample as a function of [Fe/H]. A blue dashed line at 0 km/s has been added in each panel for reference. In the metallicity range −1.2 < [Fe/H] < −0.3, all three panels exhibit a low velocity dispersion population of stars with mean values of the velocity components near 0 km/s, and dispersion in velocities of around 40-50 km/s These are the stars of the thick-disk and metal-weak thick-disk populations. The metal-weak thick-disk population likely contains stars with metallicities as least as low as [Fe/H] ~ −1.6[65], but in much lower relative proportion compared to the more metal-rich stars. At metallicities below [Fe/H] ~ −1.2, the large velocity dispersions (on the order of 100 to 150 km/s) observed in each component are associated with the broadly overlapping inner- and outer-halo populations of stars.

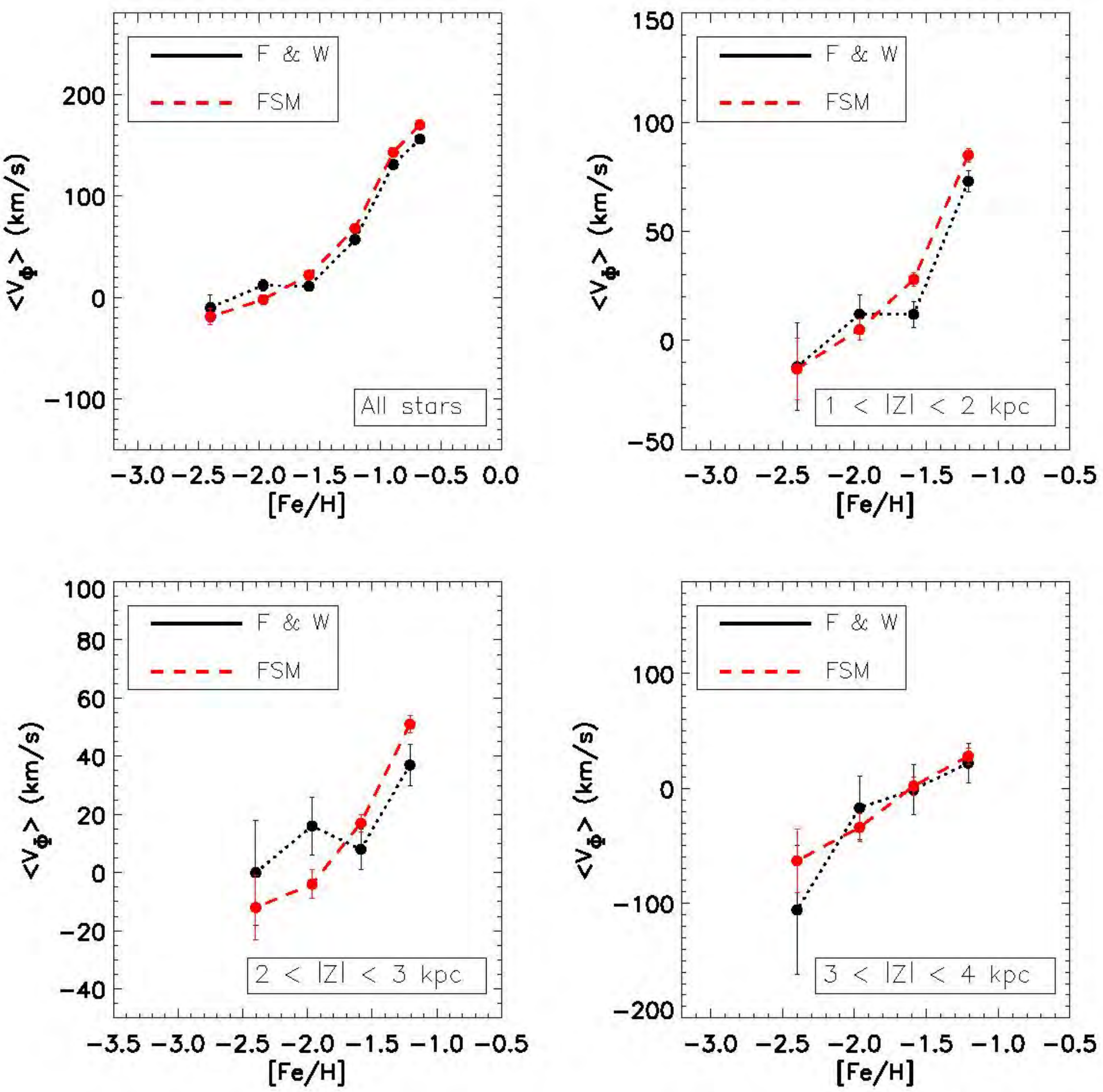

**Supplemental Figure 2 |** Comparison of the derived net rotational velocities of our sample stars with respect to the Galactocentric rest frame, $V_{\Phi}$, using two alternative methods. The black dots (and lines) correspond to the net rotations listed in Supplemental Table 2 obtained with the Frenk & White[66] analysis (F & W), which only considers the distance radial velocities. The red dots (and lines) correspond to the net rotations obtained when proper motions are employed, along with distances and radial velocities, to estimate full space motions (FSM). It is clear that the methods return equivalent estimates of net rotation when large numbers of stars are involved (upper left-hand panel). The remaining panels, which are for cuts on increasing distance from the Galactic plane, |Z|, indicate that the agreement remains acceptable, in particular at higher metallicities. Note, however, that the errors obtained from the Frenk & White analysis become quite large at lower metallicities, due to the smaller numbers of stars in the calculation.

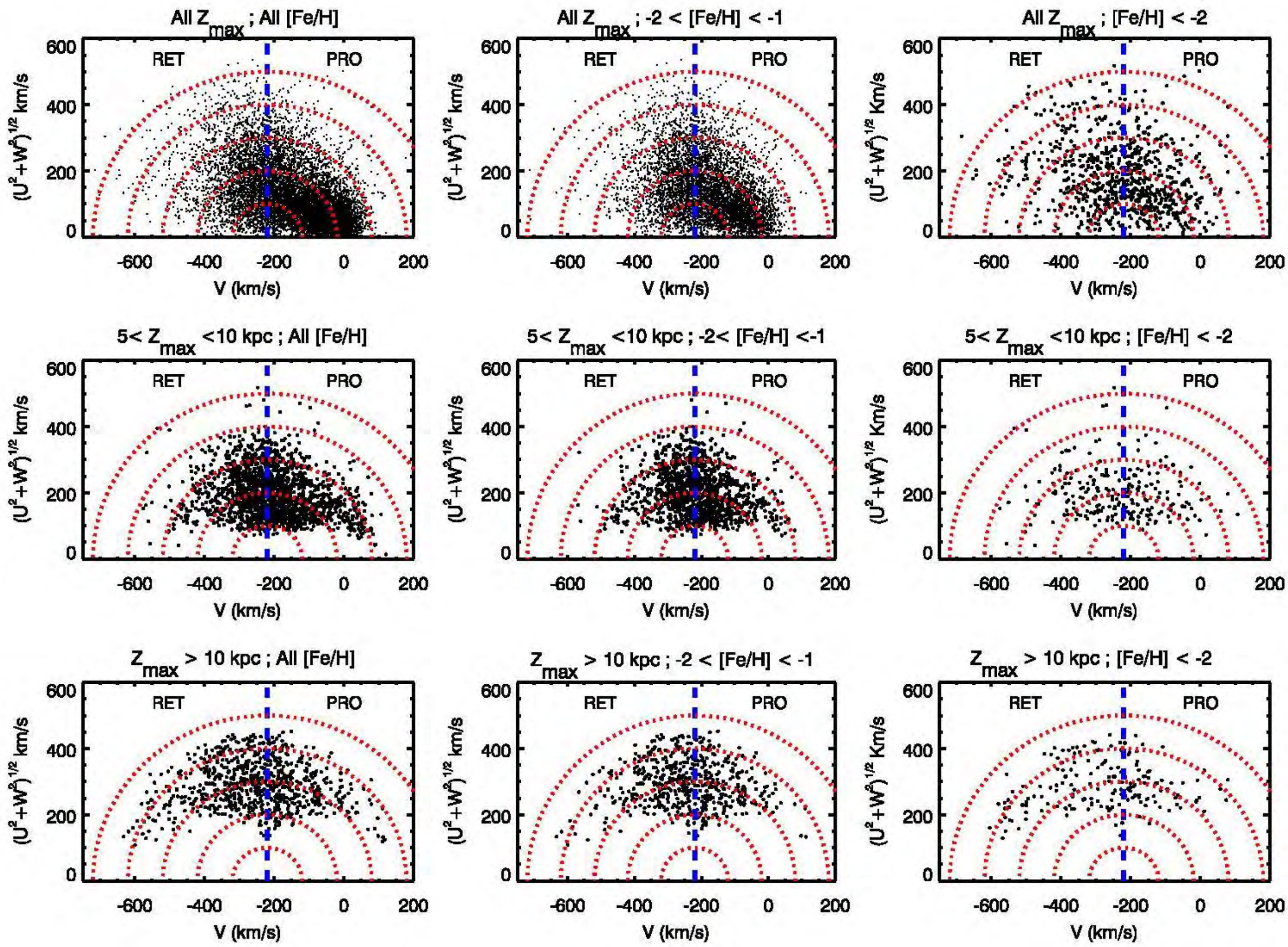

**Supplemental Figure 3 |** Toomre energy diagrams for various cuts of our sample on $Z_{max}$, the maximum distance above or below the Galactic plane that a star reaches during the course of its orbit about the Galactic centre, and on metallicity, [Fe/H]. The blue dashed line indicates the division between retrograde (RET) and prograde (PRO) orbits with respect to the Galactic centre (assuming the LSR rotates at 220 km/s[64]). The red dotted lines indicate total space velocities of the orbits, in the frame with respect to the Galactic centre; between 100 km/s (inner-most) and 500 km/s (outer-most). In the upper row, energy diagrams for stars exploring all ranges of $Z_{max}$ and various cuts in metallicity are shown – the left-most panel shows stars of all metallicities, the middle and right panels are for stars with intermediate ($-2.0 <$ [Fe/H] $< -1.0$) metallicities, and very low metallicities, [Fe/H] $< -2.0$, respectively. Inspection of this row shows the decreasing importance of stars on disk-like orbits (the highly prograde stars near 0 km/s) with declining [Fe/H]. Although a number of metal-weak thick-disk stars remain in the middle panel of this row, the majority of stars are associated with a

prograde rotation inner-halo population. In the right-hand panel of this row the metallicity selection has effectively eliminated any metal-weak thick-disk stars, and one can appreciate that the stars with retrograde orbits explore to much larger energies than the prograde stars. The orbits of the retrograde stars are also more uniformly distributed in energy than those of the prograde stars, which concentrate in the region of this diagram corresponding to lower orbital energies. The middle panels show the same cuts on metallicity but for intermediate cuts on $Z_{max}$. Even though the total numbers of stars is lower, the contrast between the higher orbital energies of the stars on retrograde orbits with the lower orbital energy stars on prograde orbits remains clear. Stars in these intervals of $Z_{max}$ are contributed from both the inner- and outer-halo populations, and exhibit different distributions in energy depending on [Fe/H]. The lower panels show the same cuts on metallicity but for stars reaching $Z_{max} > 10$ kpc, which are dominated by stars from the outer-halo population. The relative contributions of stars on various energy orbits is now much more evenly distributed between retrograde and prograde orbits, independent of [Fe/H]. The V velocities of the retrograde orbits explore to much lower V than the prograde orbits explore to higher V.

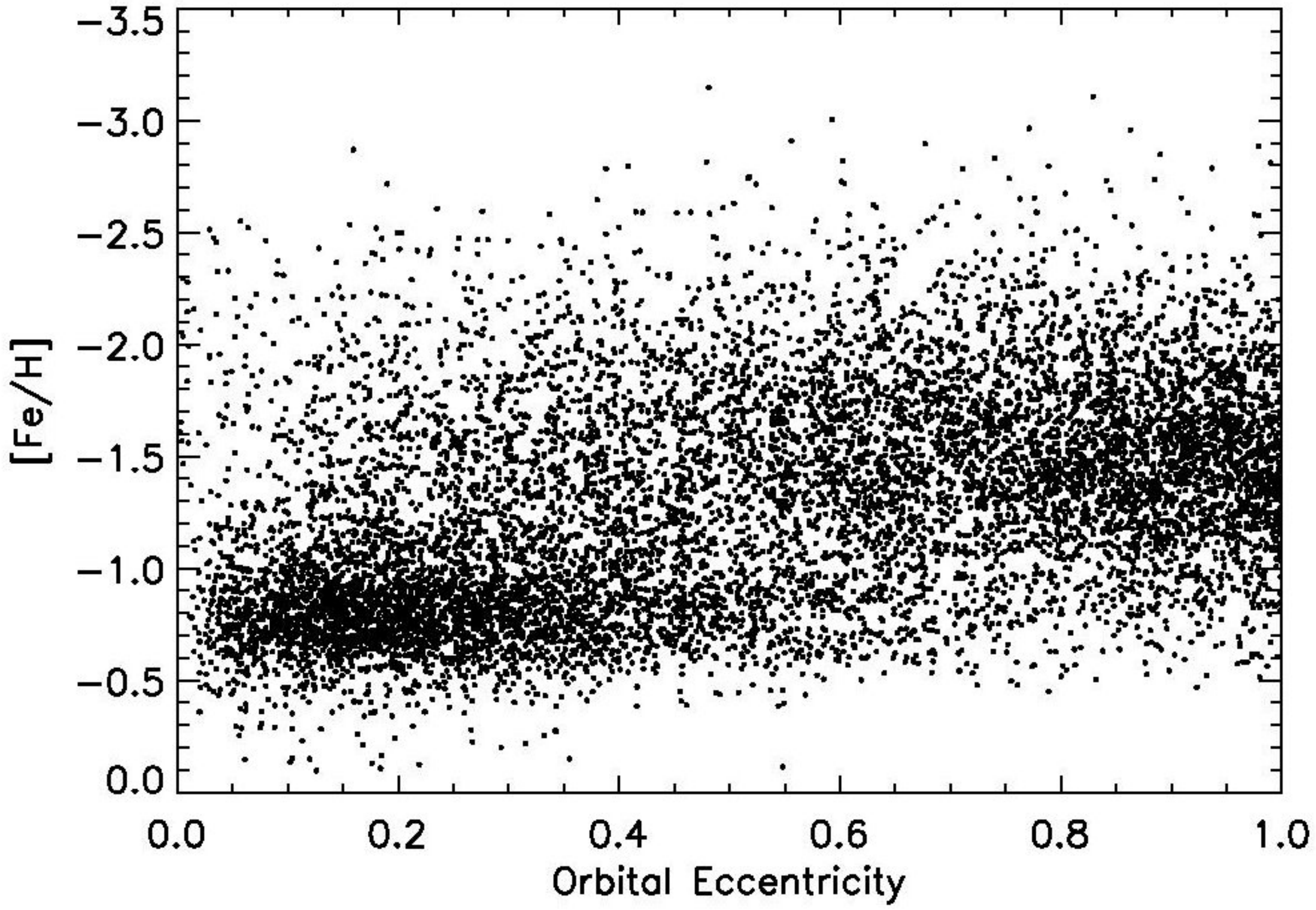

**Supplemental Figure 4 |** The relationship between [Fe/H] and orbital eccentricity for the sample of 10,123 SDSS calibration stars considered in the present analysis. At the bottom left of the panel, one can clearly distinguish the presence of moderate metallicity (−1.3 < [Fe/H] < −0.3, with a mode [Fe/H] ~ −0.8), low-eccentricity (0 to 0.4) stars of the thick-disk and metal-weak thick-disk populations. (Note that the thick disk of the Galaxy actually contains much larger numbers of stars near [Fe/H] = −0.6 than shown here; such metal-rich stars are heavily censored by the selection process used in the assembly of the calibration star sample). At the right of the panel, one can also discern the presence of low-metallicity (−2.0 < [Fe/H] < −1.0, with a mode [Fe/H] ~ −1.5), high-eccentricity (0.6 to 1.0) stars which we associate with the inner- halo population; additional members of the inner-halo population extend over all eccentricities and metallicities The outer-halo population exhibits a relatively uniform distribution of orbital eccentricities over the full range from 0.0 to 1.0, and is dominated by lower metallicity stars (−1.5 to −3.5, with a mode [Fe/H] ~ −2.0); as a result it is not easily separated from the other populations shown in this diagram.

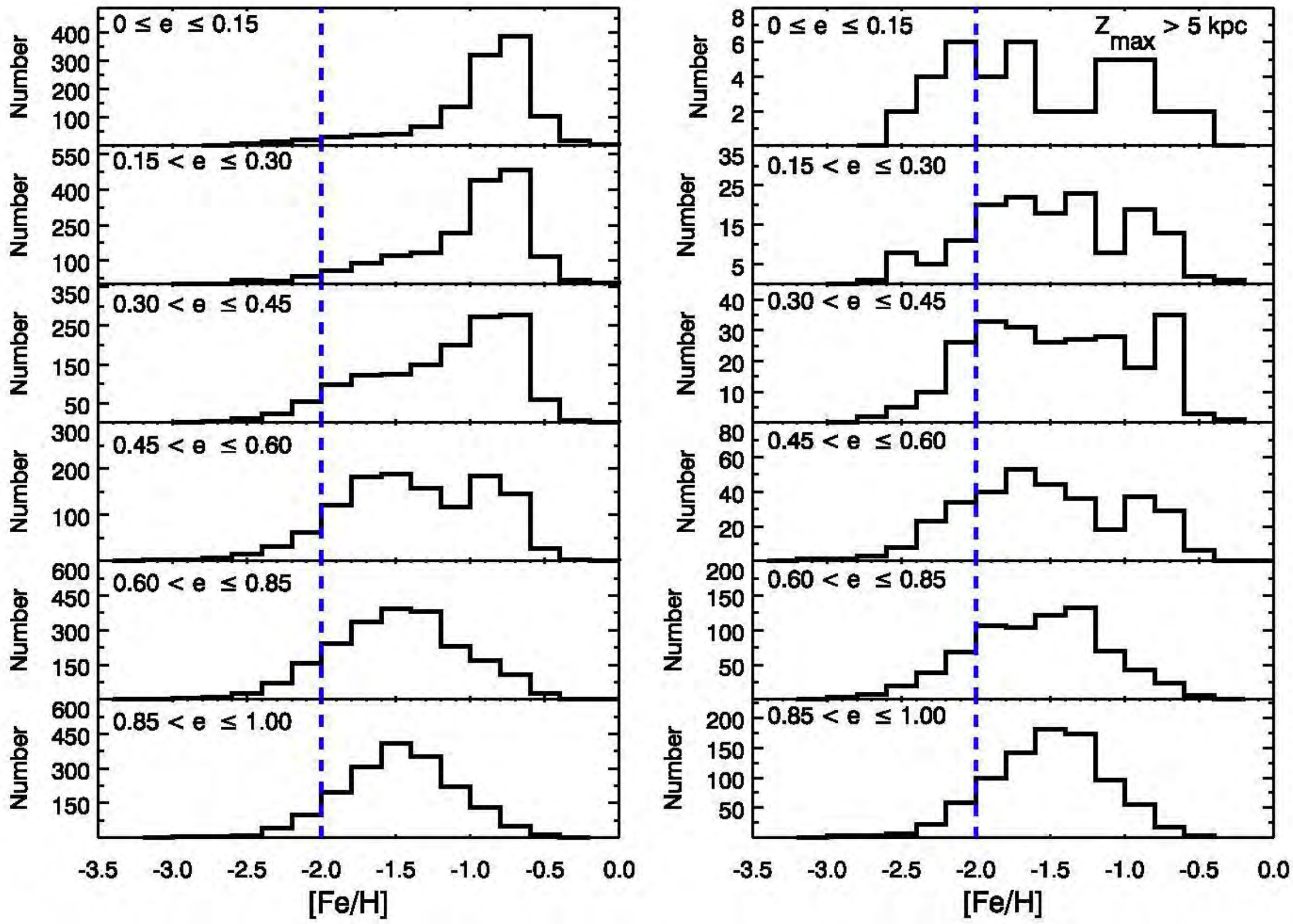

**Supplemental Figure 5 |** Distribution of [Fe/H] for various cuts in the orbital eccentricities of stars for two cuts in $Z_{max}$. A blue dashed line at [Fe/H] = −2.0 is shown for reference in both columns. In the left-hand column, all distances from the Galactic plane are considered. At low eccentricity, one clearly sees the presence of thick-disk and metal-weak thick-disk stars. As eccentricity increases, the influence of the disk-like stars decreases, and eventually, at high eccentricity, one recovers a metallicity distribution similar to the "canonical halo"; these stars are dominated by the contribution of the inner-halo population. In the right-hand column, only those stars with $Z_{max}$ > 5 kpc are shown, which effectively eliminates any stars from the disk-like populations. The behaviour at low eccentricity is clearly very different, as one sees a bimodal distribution in the upper panel, transitioning to a distribution with peak [Fe/H] ~ −1.5 at high eccentricity, but with long tails extending to lower metallicity. At intermediate eccentricities, the distribution of [Fe/H] exhibits characteristics of a mixture of populations with different metallicity distributions.

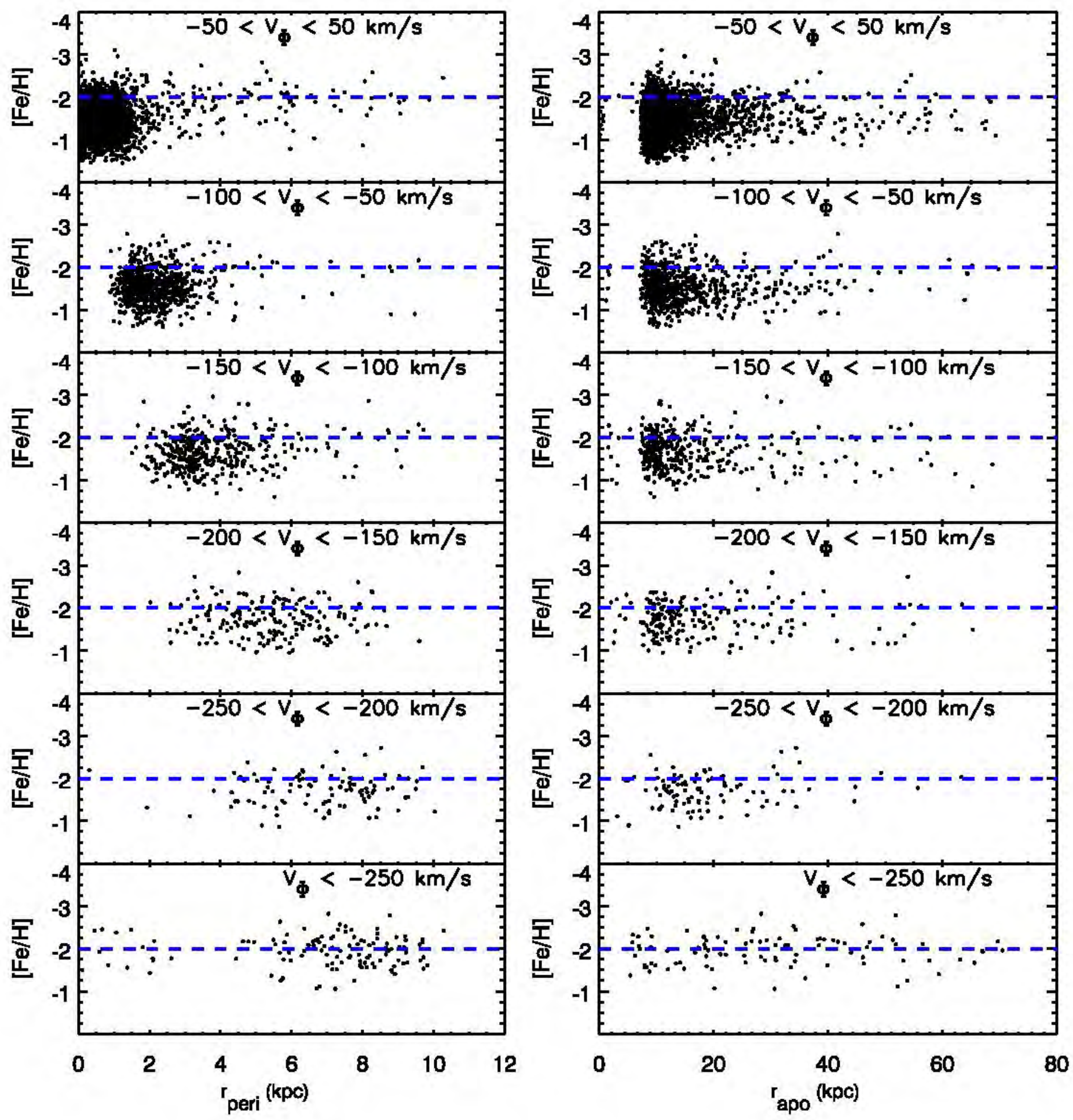

**Supplemental Figure 6 |** The nature of the derived orbits for our sample in different velocity regimes. The panels indicate metallicity as a function of the extrema of the orbits for the stars in our sample, as a function of cuts in $V_\Phi$ , from modestly prograde to highly retrograde. A blue dashed line at [Fe/H] = −2.0 is shown for reference in both columns. These diagrams illustrate the change in the nature of the orbits of the stars in the inner- and outer-halo populations over increasingly retrograde cuts in $V_\Phi$ . The left-hand column is the distribution of [Fe/H] vs. $r_{peri}$ (the distance of closest approach to the Galactic centre for a given orbit); the right-hand column is the corresponding distribution for $r_{apo}$ (the farthest distance from the Galactic centre reached for a given orbit). The upper panels of the left-hand column illustrate the dominance of inner-halo stars, which penetrate to the region near the Galactic centre (owing to their primarily high-eccentricity orbits). In this column, as one progresses to more retrograde orbits, the fraction of stars that penetrate to the Galactic centre decreases

precipitously, as does the numbers of stars with metallicity near [Fe/H] = −1.5. With only handful of exceptions, the most highly retrograde stars (which are dominated by the outer-halo population) exhibit $r_{peri} > 6$ kpc. The upper panels of the right-hand column show that the large fraction of inner-halo stars do not possess orbits that take them beyond roughly 15-20 kpc. This trend becomes more evident as the rotation velocity trends to ever more negative retrograde values. The mean metallicity of the stars on highly retrograde orbits that reach large $r_{apo}$ is clearly lower than that of the inner halo stars that reach such large distances (again, presumably due to their highly eccentric orbits) seen in the upper panels.